\documentclass[twocolumn,english,aps,pra,showpacs,amsfonts,amsmath,amssymb,superscriptaddress,footinbib,floatfix,longbibliography]{revtex4-1}
\usepackage[T1]{fontenc}
\usepackage[latin9]{inputenc}
\usepackage{mathtools}
\usepackage{amsmath}
\usepackage{amssymb}
\usepackage{graphicx}
\usepackage{lipsum}
\usepackage{siunitx}
\usepackage{braket}
\makeatletter
\usepackage{ulem}
\usepackage{color}
\usepackage{filemod}

\newcommand{\hidetxt}[1]{}

\usepackage{babel}

\makeatother

\usepackage{babel}

\usepackage{hyperref}

\begin{document}

\title{A stationary Rydberg polariton}

\author{Annika Tebben}
\email{tebben@physi.uni-heidelberg.de}
\affiliation{Physikalisches Institut, Universit\"at Heidelberg, Im Neuenheimer Feld 226, 69120 Heidelberg, Germany}

\author{Cl\'{e}ment Hainaut}
\affiliation{Physikalisches Institut, Universit\"at Heidelberg, Im Neuenheimer Feld 226, 69120 Heidelberg, Germany}

\author{Andre Salzinger}
\affiliation{Physikalisches Institut, Universit\"at Heidelberg, Im Neuenheimer Feld 226, 69120 Heidelberg, Germany}

\author{Titus Franz}
\affiliation{Physikalisches Institut, Universit\"at Heidelberg, Im Neuenheimer Feld 226, 69120 Heidelberg, Germany}

\author{Sebastian Geier}
\affiliation{Physikalisches Institut, Universit\"at Heidelberg, Im Neuenheimer Feld 226, 69120 Heidelberg, Germany}

\author{Gerhard Z\"urn}
\affiliation{Physikalisches Institut, Universit\"at Heidelberg, Im Neuenheimer Feld 226, 69120 Heidelberg, Germany}

\author{Matthias Weidem\"uller}
\email{weidemueller@uni-heidelberg.de}
\affiliation{Physikalisches Institut, Universit\"at Heidelberg, Im Neuenheimer Feld 226, 69120 Heidelberg, Germany}

\begin{abstract}
We propose a novel scheme for coupling a Rydberg state to a stationary light polariton, based on a dual-V level scheme. We investigate the properties of the resulting stationary Rydberg polariton, and show that its form and its quadratic dispersion relation closely resemble that of the stationary light polariton of the underlying dual-V scheme. We consider the influence of a Rydberg impurity on the system and find strong interaction-induced absorption of the involved probe field. The proposed scheme for a stationary Rydberg polariton might find applications for realizing interacting polaritons with increased interaction time.

\end{abstract}

\date{\today}

\maketitle
 
\section{Introduction}
\label{sec:intro}
Interactions between polaritons that are composed of light and matter open the possibility for investigating many-body physics with photons and exotic states of light \cite{Moos:manyBodyRydPolaritons:PRA15,Otterbach:PhotonCrystal:PRL13,Carusotto:QuantumFluid:RMP13,Bienias:ScatteringResonances:PRA2014,Chang:Review:NatPhot14}. The combination of electromagnetically induced transparency (EIT) with the exceptional strong interactions between Rydberg atoms has enabled the realization of interacting Rydberg polaritons \cite{Gorshkov:PhotIntBlockade:PRL11,Petrosyan:EITRydAtoms:PRL11,Peyronel:QuantREIT:Nature12,Firstenberg:AttractivePhot:Nature13,Busche:Contactless:NatPhys17,Thompson:SymmIntPhotons:Nature17,Liang:ThreePhotBound:Science18,Stiesdal:ThreeBody:PRL18,Cantu:RepulsivePhotons:Nature2020} and quantum optics applications \cite{Chaneliere:QuantMemExp:Nature05,Dudin:PhotonAntibunch:Science12,Baur:PhotonSwitch:PRL14,Gorniaczyk:PhotonTransistor:PRL14,Tiarks:PhotonTransistor:PRL14,Tiarks:PiShift:Science16,Murray:PhotonSubstraction:PRL18,Tiarks:PhotonGate:NatPhys19,Ornelas:SinglePhotonSource:Optica20}.
However, photon-photon interactions implemented with this approach are limited by the restricted interaction time between the propagating polaritons \cite{Otterbach:PhotonCrystal:PRL13,Kim:RydPolaritonThermalization:CommPhys21}.

Storing light as an atomic spin wave in the Rydberg EIT medium \cite{Fleischhauer:DarkStatePolariton:PRL00,Fleischhauer:QuantumMemory:PRA02,Distante:StorageEnhanced:PRL16} has been demonstrated to enhance the effect of interactions \cite{Distante:StorageEnhanced:PRL16,Busche:Contactless:NatPhys17} due to an increased interaction time. Inspired by this idea enabling interactions between stationary photonic excitations could increase the interaction time for two photons compared to the propagating case. This could for example be achieved by coupling a stationary light polariton (SLP) \cite{Andre:proposalSLP:PRL02,Bajcsy:SLP:Nature03,Andre:SLP:PRL05,Zimmer:SLPlambdaScheme:OptCom06,Zimmer:SLPdualV:PRA08,Lin:SLPcoldAtoms:PRL09,Park:quantumSLP:PRX18,Everett:StationaryLIght:AdvQuantTech19} to a Rydberg state.
One approach in this direction included to use the second meta-stable state in a diamond like level scheme as a Rydberg state \cite{Nikoghosyan:RydbergSPL:PRA12,Pimenta:RydSLP:ArXiv18}. For this scheme Bose-Einstein condensation in the presence of Rydberg interactions has been proposed \cite{Nikoghosyan:RydbergSPL:PRA12}.  
Moreover, an implementation of SLPs based on a dual-V level scheme \cite{Fleischhauer:SLPBEC:PRL08,Zimmer:SLPdualV:PRA08} has been exploited in conjunction with Rydberg interactions for coherent switching between slow- and stationary light polaritons \cite{Murray:PolaritonSwitching:PRX17}.

Here, we propose a novel level scheme for coupling a Rydberg state to a stationary light polariton. We base our proposal on the dual-V level \cite{Zimmer:SLPdualV:PRA08} scheme as it features similar wavelengths of the probe and control fields. This drastically reduces experimental issues resulting from phase matching compared to the diamond-like level scheme. Indeed, it is known that a phase mismatch reduces the lifetime of the SLP in experiments \cite{Chen:TwoStoppedLight:PRL12}, which would ultimately restrict the possible observation and utilization of interaction effects.

This paper is structured as follows: First, we present in Sec.~\ref{sec:dressing} our level scheme that couples a Rydberg state to a stationary light polariton. Based on the discussion in ref.~\cite{Zimmer:SLPdualV:PRA08}, we explicitly verify that our system supports a stationary photonic excitation with a quadratic dispersion relation, which we call a stationary Rydberg polariton. In order to learn about how interactions originating from the Rydberg character affect the proposed scheme, we investigate the effect of a Rydberg impurity on the system in Sec.~\ref{sec:interaction}. By solving the propagation equation for the probe field modes, we examine the transmission, reflection and absorption properties of the considered system, which could be accessed experimentally. Finally, we conclude in Sec.~\ref{sec:conclusion} and given an outlook on possible applications of the scheme.

\begin{figure}[t!]
	\includegraphics[width=0.7\linewidth]{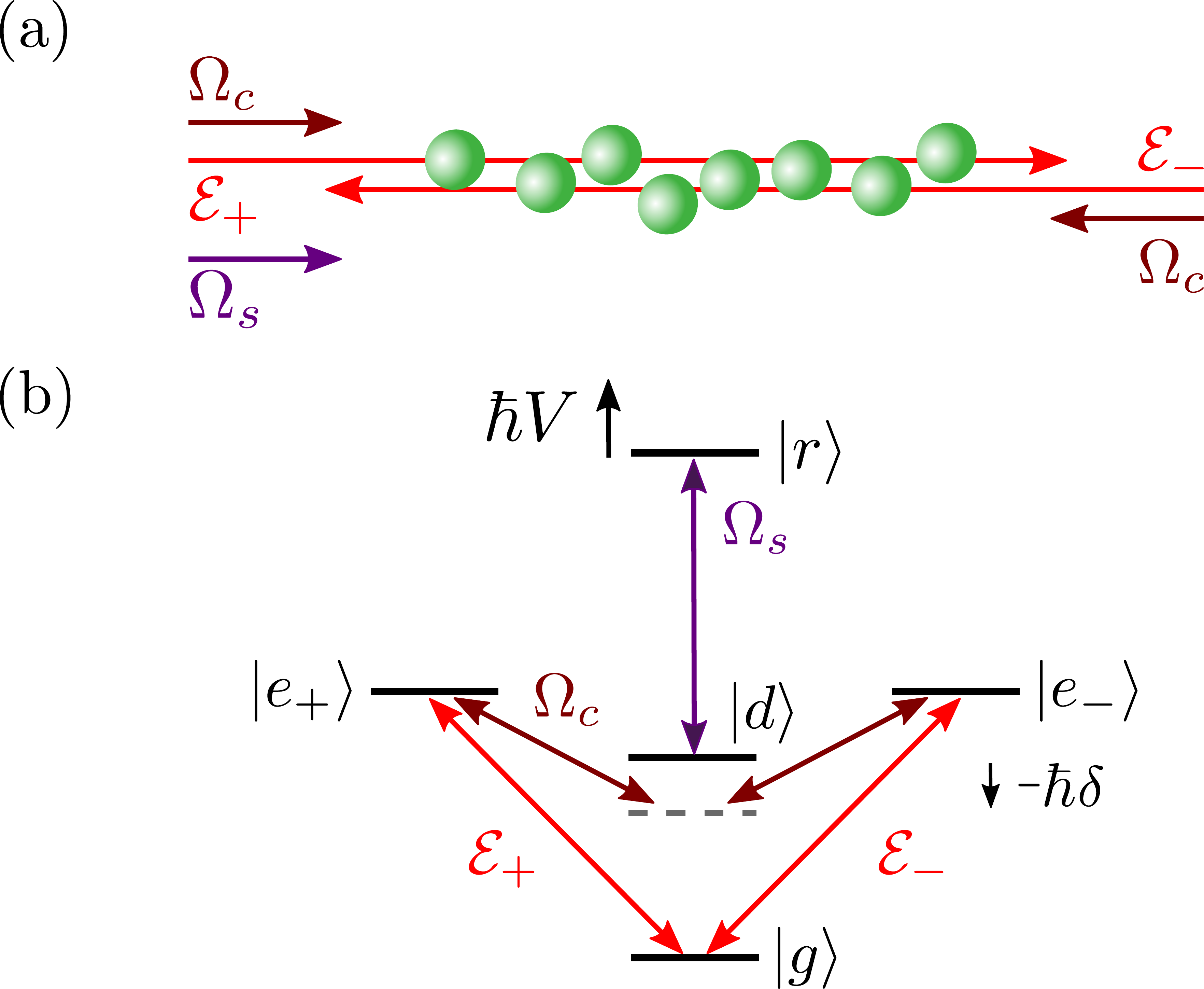}
	\caption{(a) Schematic illustration of the applied fields for the creation of a stationary Rydberg polariton. 
		(b) Level scheme for a stationary Rydberg polariton based on a dual-V level scheme. 
		In the presence of a Rydberg impurity, the Rydberg state experiences an energy shift $\hbar V$. For details see the main text.}
	\label{fig:level-schemes} 
\end{figure}

\newpage
\section{Realization of a stationary Rydberg polariton}
\label{sec:dressing}

We consider a dual-V scheme realization of a SLP \cite{Fleischhauer:SLPBEC:PRL08,Zimmer:SLPdualV:PRA08}, as schematically depicted in Fig.~\ref{fig:level-schemes}(a). Here, two counter-propagating classical control fields with equal Rabi frequency $\Omega_c$ induce stationary light conditions for two quantized probe modes $\hat{\mathcal{E}}_\pm$. The underlying atomic level scheme is shown in Fig.~\ref{fig:level-schemes}(b) and is composed of a ground-state $\ket{g}$, a meta-stable state $\ket{d}$ and two intermediate states $\ket{e_\pm}$ that decay with rate $\gamma$. We couple the meta-stable state $\ket{d}$ to a Rydberg state $\ket{r}$ using a classical Rydberg coupling field with Rabi frequency $\Omega_s$.

The equations of motion for the probe field amplitudes and for the continuous bosonic field operators $\hat{D}$, $\hat{S}$ and $\hat{P}_\pm$ of the coherences \cite{Fleischhauer:DarkStatePolariton:PRL00,Fleischhauer:QuantumMemory:PRA02} in the system read in analogy to refs. \cite{Murray:PolaritonSwitching:PRX17,Zimmer:SLPdualV:PRA08,Fleischhauer:SLPBEC:PRL08}
\begin{align}
\partial_t \hat{\mathcal{E}}_\pm &=\mp c\partial_z\hat{\mathcal{E}}_\pm-i G\hat{P}_\pm\, ,\label{equ:EOM_E}\\
\partial_t \hat{P}_\pm &= -i G\hat{\mathcal{E}}_\pm-i \Omega_c \hat{D}-\bar{\gamma} \hat{P}_\pm\, ,\\
\partial_t \hat{D}&=-i\Omega_c (\hat{P}_++\hat{P}_-)-i\Omega_s\hat{S}-i\delta\hat{D}\, ,\\
\partial_t \hat{S}&=-i\Omega_s\hat{D}-i (\Delta_s+\delta)\hat{S}\, .
\label{equ:EOM_S}
\end{align}
Here, we have defined the one-dimensional atomic density $\rho$ and have used the short notation $\hat{\mathcal{O}}(z,t)=\hat{\mathcal{O}}$ for all operators. Furthermore, we have introduced the two-photon detuning $\delta$, which is considered to be the same for the two ground- to meta-stable state transitions, and the detuning $\Delta_s$ of the Rydberg coupling field. $G=g\sqrt{\rho}$ with the single-atom coupling strength $g$, $\bar{\gamma}=\gamma/2$ and $c$ is the speed of light. One can cast the equations of motion into the form
\begin{equation}
i\partial_t\Upsilon=\mathcal{H}_\text{eff}\Upsilon\, ,
\label{equ:EOM_matrix}
\end{equation}
where the column vector $\Upsilon=\left(\hat{\mathcal{E}}_+,\hat{\mathcal{E}}_-,\hat{D},\hat{S},\hat{P}_+,\hat{P}_-\right)^T$ has been defined. The coefficient matrix $\mathcal{H}_\text{eff}$ can be readily obtained from Eqs.~(\ref{equ:EOM_E}) to (\ref{equ:EOM_S}).

In the following, we assume a resonant coupling to the Rydberg state ($\Delta_s=0$). We observe that diagonalizing the subsystem $\{\hat{D},\hat{S}\}$  yields two eigenstates, which have eigenenergies $-\hbar\delta\pm\hbar\Omega_s$, respectively. Under the condition $\delta=\pm\Omega_s$ one of them is on two-photon resonance with the ground state, while the other one is shifted in energy by $|2\hbar\Omega_s|$. Due to the resulting similarity with the dual-V scheme without the Rydberg state, which has been studied in detail in ref.~\cite{Zimmer:SLPdualV:PRA08}, we expect that the system can support a stationary polariton similar to the SLP in the dual-V scheme.

In order to confirm this expectation, we follow the ideas of ref.~\cite{Zimmer:SLPdualV:PRA08} and investigate $\mathcal{H}_\text{eff}$ in momentum space. Considering the case $\delta=\Omega_s$ \footnote{The case $\delta=-\Omega_s$ can be treated analogously and also leads to one stationary Rydberg polariton in the system.}
and diagonalizing $\mathcal{H}_\text{eff}$ under this condition, we obtain a unique dark-state of the form
\begin{equation}
\hat{\Psi}=\frac{1}{\mathcal{N}}\left[ \Omega_c \left(\hat{\mathcal{E}}_++\hat{\mathcal{E}}_-\right)-G\left(\hat{D}+\hat{S}\right) \right]\, ,
\label{equ:SLP_dark-state_B}
\end{equation}
with $\mathcal{N}=\left[G^2+2\Omega_c^2\right]^{1/2}$. It has exactly the same form as the stationary light polariton in the dual-V scheme \cite{Zimmer:SLPdualV:PRA08} with the replacement $\hat{D}\rightarrow\hat{D}+\hat{S}$.
\begin{figure}[t!]
	\includegraphics[width=0.85\linewidth]{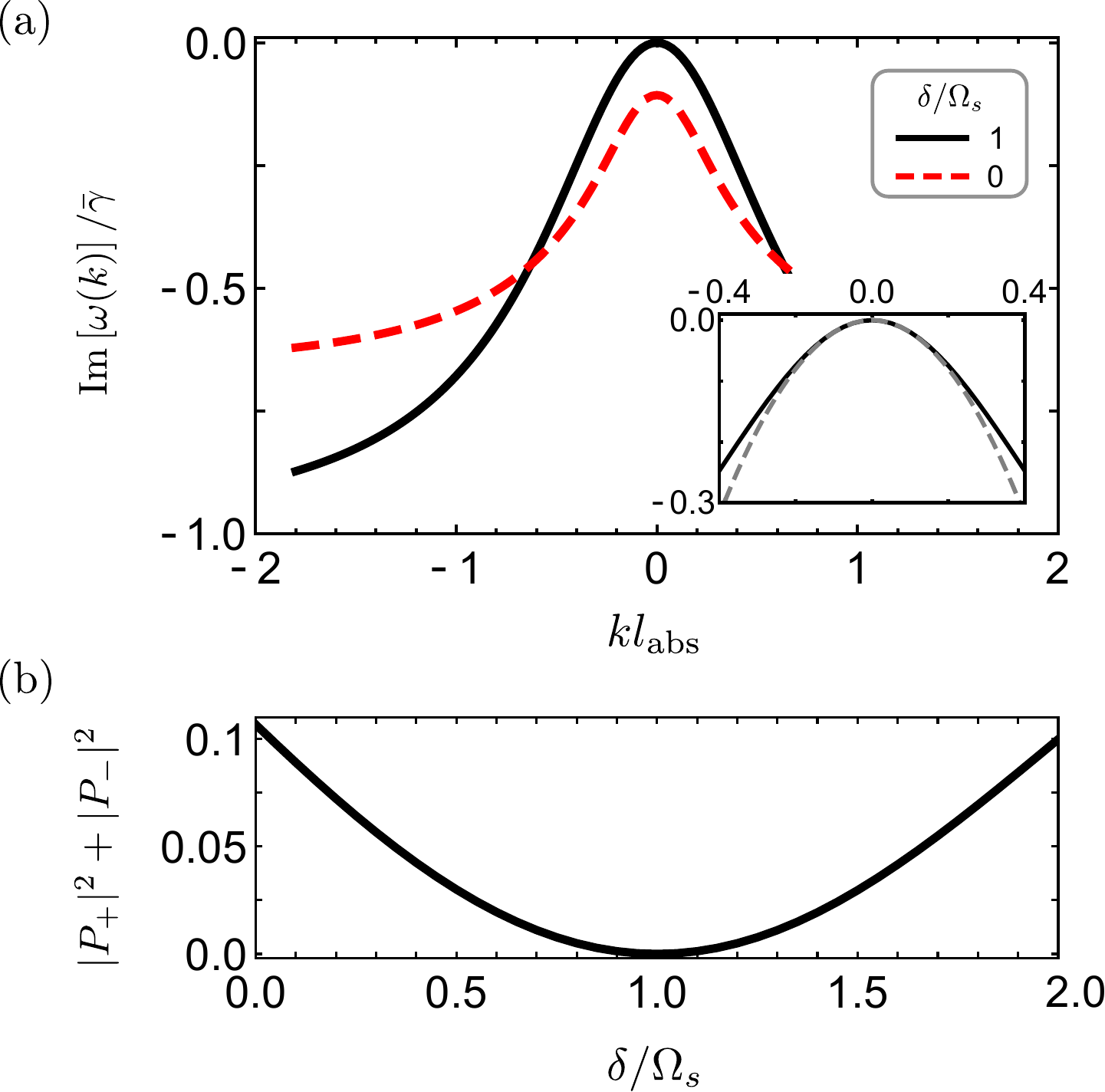}
	\caption{(a) Dispersion relation $\text{Im}\left[\omega(k)\right]/\bar{\gamma}$ as a function of the photon momentum $k$ scaled by the absorption length $l_\text{abs}=c\bar{\gamma}/G^2$ \cite{Murray:PolaritonSwitching:PRX17}. The inset shows and enlargement of the plot for small photon momenta together with the approximate dispersion relation (gray dashed line) given in Eq.~(\ref{equ:dispersion_relation}). (b) Probability $|P_+|^2+|P_-|^2$ to be in one of the intermediate states as a function of $\delta/\Omega_s$ for $k=0$. Parameters for both plots are
		$\Omega_s/\gamma=\Omega_c/\gamma=G/\gamma=1$.}
	\label{fig:polariton-analysis} 
\end{figure}
In order to obtain the dispersion relation associated to this dark-state, we follow the calculation in ref.~\cite{Zimmer:SLPdualV:PRA08}. For small photon momenta and up to second order, the dispersion relation reads
\begin{equation}
\omega(k)\approx -\frac{ic^2\bar{\gamma}\Omega_c^2}{G^2(G^2+\Omega_c^2)}k^2\, .
\label{equ:dispersion_relation}
\end{equation}
In the inset of Fig.~\ref{fig:polariton-analysis}(a) we compare this result with a numeric calculation (black solid line) and find good agreement.
We emphasize that with the replacement $\Omega_c\rightarrow\Omega_c/\sqrt{2}$  Eq.~(\ref{equ:dispersion_relation}) is identical to the dispersion relation that has been found for the dual-V scheme without a Rydberg component \cite{Zimmer:SLPdualV:PRA08}. The associated effective mass of the stationary polariton can be obtained from Eq.~(\ref{equ:dispersion_relation}). Interestingly, it can be controlled with a particular choice of $\Omega_c$, but does not depend on $\Omega_s$. From the above discussion we conclude that our proposed scheme supports a stationary light polariton with a strong Rydberg component, which we call a stationary Rydberg polariton. 

The characteristics of a stationary photonic excitation are no contribution from the intermediate states, equal contributions of the two probe field modes and a quadratic dispersion relation. We want to underline that these characteristics can be used to investigate under which conditions the stationary Rydberg polariton is supported in the medium. In the previous discussion we focused on the particular situation where $\delta=\Omega_s$. In Fig.~\ref{fig:polariton-analysis}(a), we also show the dispersion relation of the same eigenstate of $\mathcal{H}_\text{eff}$ for $\delta=0$. It also displays a quadratic behavior for small photon momenta. Nevertheless, the state possesses a non-zero probability to be in one of the intermediate states, as shown in Fig.~\ref{fig:polariton-analysis}(b). This means that the conditions for having a stationary Rydberg polariton in the medium are broken. Indeed, we find that this also holds for all other ratios $\delta/\Omega_s\neq 1$, as shown in Fig.~\ref{fig:polariton-analysis}(b). Thus, already a small deviation from the condition $\delta=\Omega_s$, where one of the eigenstates of the subsystem $\{\hat{D},\hat{S}\}$ is resonantly coupled, breaks the condition for the stationary Rydberg polariton. This is in accordance with the requirement to obey the EIT resonance.

\section{Transmission and reflection properties in the presence of a Rydberg impurity}
\label{sec:interaction}
Having outlined under which conditions a stationary Rydberg polariton is supported in our scheme, we now investigate how the interaction with a Rydberg impurity affects the system. In the simplest case, the interaction with the impurity can be modeled as a level shift $\hbar V$ of the Rydberg level, which can be captured by adding the term $iV\hat{S}$ in Eq.~(\ref{equ:EOM_S}) \cite{Murray:PolaritonSwitching:PRX17}. In the previous section we found that only if $|\delta|=\Omega_s$ holds, the conditions for a stationary Rydberg polariton are met. Therefore, based on this simple picture, we expect that the conditions for the stationary Rydberg polariton are altered or even broken in the presence of a Rydberg impurity. 

\begin{figure}[t!]
	\includegraphics[width=0.85\linewidth]{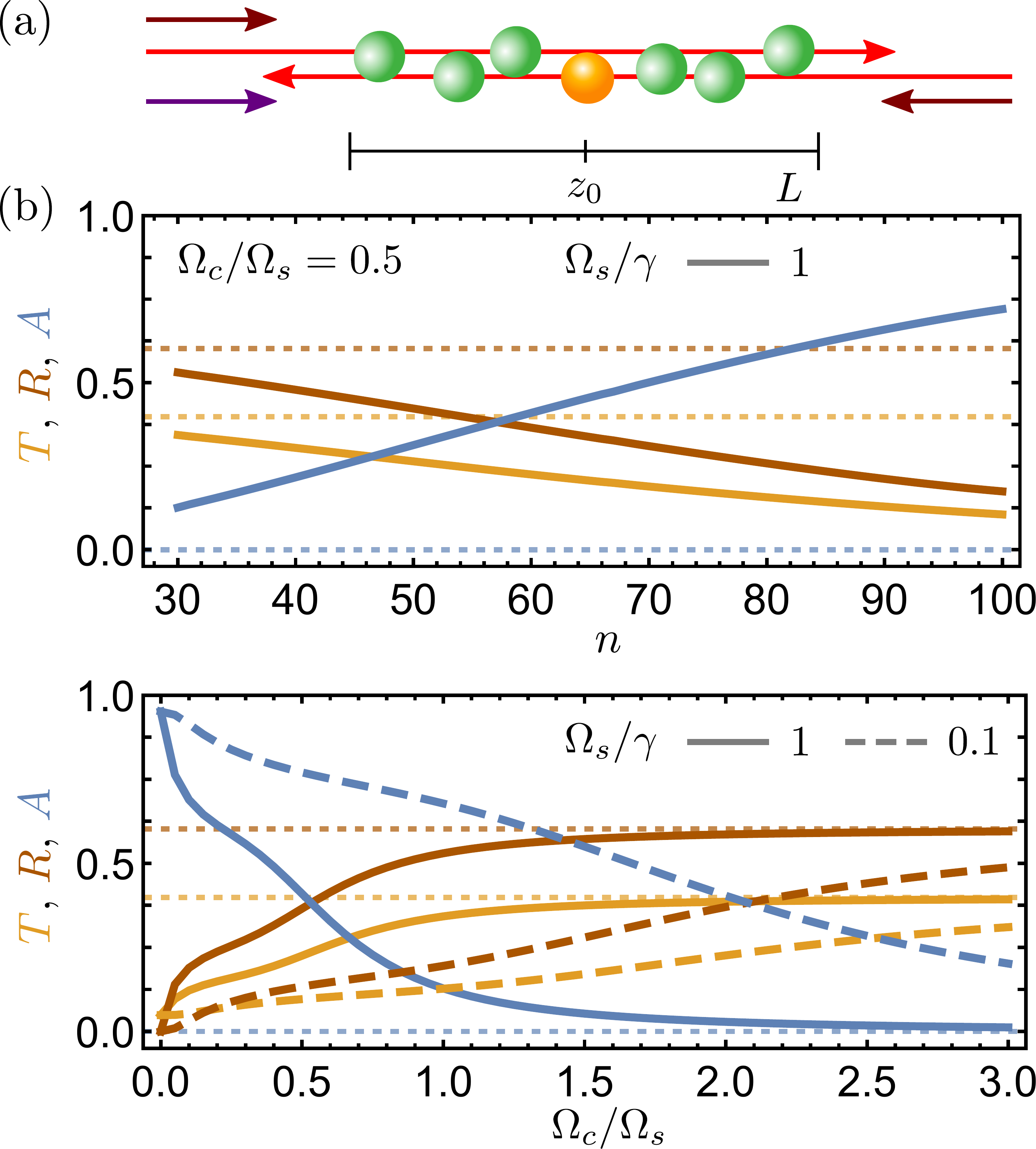}
	\caption{ (a) Schematic illustration of a Rydberg impurity located at position $z_0$ in a medium of length $L$. (b) Transmission $T$, reflection $R$ and absorption $A$ coefficients of the probe field as a function of the Rydberg quantum number $n$ (upper graph) and the ratio $\Omega_c/\Omega_s$ for $n=60$ (lower graph) for the parameters as indicated in the figure. Other parameters are $\delta/\Omega_s=1$, $\Delta_s=0$, $L=\SI{40}{\mu m}$ and an optical depth of $3$. Moreover, we consider coupling and interaction strengths for the realization with $^{87}$Rb atoms as discussed in App.~\ref{app:exp}. Horizontal dotted lines show the expected values for the reflection, transmission and absorption coefficients (from top to bottom) in the absence of the impurity, respectively.}
	\label{fig:transmission} 
\end{figure}

For the following investigation, we consider a specific situation where a Rydberg impurity is positioned at $z_0$ in a medium of length $L$, as schematically depicted in Fig.~\ref{fig:transmission}(a). This Rydberg impurity is assumed to be decoupled from the applied fields and to interact with the state $\ket{r}$ via van der Waals interactions $V_\text{vdW}=V(z-z_0)=C_6/(z-z_0)^6$. Furthermore, we assume that the probe field is incident from the left onto the medium under continuous wave (cw) conditions. This allows us to treat the equation of motion, Eq.~(\ref{equ:EOM_matrix}), in the stationary limit. It is known that if the conditions for a stationary Rydberg polariton in the medium are met, the probe field will get reflected from the medium with a certain probability \cite{Bajcsy:SLP:Nature03,Murray:PhotonSubstraction:PRL18}. However in the presence of the Rydberg impurity, we expect a finite probability to be in the intermediate states and thus partial absorption of the probe field.

In order to derive the transmission, reflection and absorption coefficients we solve the equations of motion, Eq.~\ref{equ:EOM_matrix}. For this purpose, we follow the calculations performed in ref. \cite{Murray:PolaritonSwitching:PRX17} and obtain in the steady-state limit the propagation equation 
\begin{equation}
i\partial_z\mathbf{E}(z)=\mathbf{M}(z)\mathbf{E}(z)
\label{equ:prop-equation}
\end{equation}
for the amplitudes $\mathcal{E}_\pm$ of the probe fields. Here, $\mathbf{E}(z)=\left\{\mathcal{E}_+(z),\mathcal{E}_-(z)\right\}$. The propagation matrix $\mathbf{M}(z)$ is given by
\begin{equation}
\mathbf{M}(z)=\left( \begin{array}{cc}
\chi_{++}(z)&\chi_{+-}(z)\\
-\chi_{+-}(z)&-\chi_{++}(z)\\
\end{array}\right)\, .
\end{equation}
Here,
\begin{align}
\chi_{++}(z)&=\frac{-iG^2}{c\bar{\gamma}}\frac{(V_\text{vdW}-\delta_r)(\bar{\gamma}\delta-i\Omega_c^2)+\bar{\gamma}\Omega_s^2}{(V_\text{vdW}-\delta_r)(\bar{\gamma}\delta-2i\Omega_c^2)+\bar{\gamma}\Omega_s^2}\, ,\\
\chi_{+-}(z)&=\frac{G^2}{c\bar{\gamma}}\frac{(V_\text{vdW}-\delta_r)\Omega_c^2}{(V_\text{vdW}-\delta_r)(\bar{\gamma}\delta-2i\Omega_c^2)+\bar{\gamma}\Omega_s^2}
\end{align}
with $\delta_r=\delta+\Delta_s$,
describe effective potentials for the right- and left-propagating probe fields and explicitly show the coupling between the two modes. As we consider a situation where a probe field is incident from the left onto the medium, we solve Eq.~(\ref{equ:prop-equation}) with the boundary conditions $\mathcal{E}_+(z=0)=\mathcal{E}_0$ and $\mathcal{E}_-(z=L)= 0$. From the solutions for the probe field amplitudes $\mathcal{E}_\pm$ we compute the transmission and reflection coefficients defined as
\begin{equation}
T=\frac{|\mathcal{E}_+(z=L)|}{|\mathcal{E}_0|}\quad \text{and}\quad R=\frac{|\mathcal{E}_-(z=0)|}{|\mathcal{E}_0|}
\end{equation}
respectively, as well as the absorption coefficient $A=1-T-R$. First, we investigate how theses quantities change, when interactions are gradually increased. This can be done by varying the quantum number $n$ of the Rydberg state in which the Rydberg impurity resides.
The result is shown in the upper graph of Fig.~\ref{fig:transmission}(b). For increasing interaction strength, absorption of the probe field increases at the cost of both transmission and reflection probabilities. This is due to the fact that stronger interactions imply a larger level shift. As a consequence the probability to be the intermediate states and thus absorption of the probe field are increased.

In the lower graph of Fig.~\ref{fig:transmission}(b) we explore for $n=60$ how the amount of interaction-induced absorption depends on system parameters such as the Rabi frequency of the control field and the Rydberg coupling strength. For $\Omega_c/\Omega_s =0$, the system reduces to a two-level system with transmission and absorption coefficients given by the optical depth of the medium. 
For $\Omega_c/\Omega_s \gg 1$, absorption goes to zero, while the transmission and reflection coefficients approach the values of the corresponding non-interacting system under stationary Rydberg polariton conditions. This is due to the fact that the EIT linewidth $\Omega_c/\gamma$ is large compared to the interaction strength, such that the system becomes insensitive to interactions. For small $\Omega_c/\Omega_s$ the situation is reversed. Here, the EIT linewidth is small such the system is very sensitive to interactions and displays strong absorption of the probe field. Finally, for a given ratio $\Omega/\Omega_s$ we observe that the effect of an interaction induced energy shift is increased for smaller ratios of $\Omega_s/\gamma$. This is a consequence of an increased atomic part of the polariton and thus a larger probability to be in the Rydberg state.

\section{Conclusion and Outlook}
\label{sec:conclusion}
In this work we have proposed to endow a stationary light polariton with a Rydberg character by coupling one of the ground states of a dual-V scheme to a Rydberg state. We have shown that a stationary Rydberg polariton under resonant coupling to the Rydberg state ($\Delta_s=0$) is supported in the medium if the detuning of the control fields is chosen appropriately as $|\delta|=\Omega_s$. This polariton is directly linked to the stationary light polariton of the underlying dual-V scheme with a quadratic dispersion relation \cite{Zimmer:SLPdualV:PRA08}.
Moreover, we have shown that the presence of the Rydberg impurity changes the transmission, reflection and absorption properties for a probe field incident on a medium under stationary Rydberg conditions. In particular, the presence of the impurity leads to a decreased transmitted and reflected probe signal, and the amount of interaction-induced absorption can be tuned by the ratio $\Omega_c/\Omega_s$. This feature could be exploited to probe the presence of the Rydberg impurity in the medium, similar to the method of interaction enhanced imaging \cite{Guenter:IEI:Science13}.

In this work, we have focused on the simplest case of an interaction-induced energy shift in the presence of a Rydberg impurity probed by continuous wave fields. Creating the stationary Rydberg polaritons inside the atomic medium instead, e.g. by retrieving a stored spin wave by simultaneously turning on the two control fields \cite{Andre:proposalSLP:PRL02,Bajcsy:SLP:Nature03}, would allow to study interactions between the polaritons themselves. In this case, we expect that interactions in the order of $\Omega_s$ between the polaritons result from the resonant coupling to the Rydberg state. Moreover, the interaction time could be increased compared to the situation of propagating polaritons, but will be limited by the lifetime of the polaritons \cite{Zimmer:SLPlambdaScheme:OptCom06}. Thus, extending our work to interactions between two or more stationary Rydberg polaritons would thus be interesting in future work.

A theoretical treatment beyond two polaritons requires to take into account correlations between the quasi-particles, which is challenging for a strongly interacting many-body system. In order to provide an alternative approach, we want to point out that in our level scheme the Rydberg state can also be coupled under Rydberg dressing conditions \cite{Johnson:RydDressing:PRA10, Balewski:RydDressingExp:NJP14, Jau:RydDressingEntangling:Nature16,Arias:RydDressedInterfereometer:PRL19}, such that a perturbative treatment of interactions becomes possible. Under Rydberg dressing conditions, meaning $|\Delta_s|\gg\Omega_s$ and $\delta=0$, the dark-state has the form 
\begin{equation}
\hat{\Psi}'=\frac{1}{\mathcal{N'}}\left[ \Omega_c \left(\hat{\mathcal{E}}_++\hat{\mathcal{E}}_-\right)-G\left(\hat{D}+\frac{\Omega_s}{\Delta_s}\hat{S} \right)\right]
\label{equ:SLP_dark-state_A}
\end{equation}
with normalization ${\mathcal{N}'=\left[G^2(1+\Omega_s^2/|\Delta_s|^2)+2\Omega_c^2\right]^{1/2}}$. Using the same analysis as in Sec.~\ref{sec:dressing} we can show, that also in this case our scheme supports a stationary Rydberg polariton, which follows the dispersion relation of the genuine ground-state dual-V SLP \cite{Zimmer:SLPdualV:PRA08,Murray:PolaritonSwitching:PRX17}. The Rydberg component of the polariton is inherently small with $\Omega_s/|\Delta_s|\ll 1$ and can be tuned by this ratio. This permits to treat interactions between the stationary Rydberg polaritons themselves perturbatively. The previous discussion suggests that our work provides a route towards investigating polariton interactions with increased interaction time, due to the stationary nature of the stationary Rydberg polariton.

\subsection*{ACKNOWLEDGMENTS}
The authors thank Thomas Pohl for many stimulating discussions and for comments on the manuscript, and Michael Fleischhauer for sharing his insights into the dual-V stationary light polariton.
This work is part of and supported by the DFG Priority Program "GiRyd 1929" (DFG WE2661/12-1) and the Heidelberg Center for Quantum Dynamics, and it is funded by the Deutsche Forschungsgemeinschaft (DFG, German Research Foundation, Project-ID 273811115, SFB 1225 ISOQUANT). A.T. acknowledges support from the Heidelberg Graduate School for Fundamental Physics (HGSFP). C.H. acknowledges support from the Alexander von Humboldt foundation.

\appendix

\section{Considerations for an experimental implementation}
\label{app:exp}
The level scheme presented in Fig.~\ref{fig:level-schemes}(b) can for example be realized in a gas of $^{87}$Rb atoms, similar to ref. \cite{Murray:PolaritonSwitching:PRX17}. Here, the two ground states $\ket{g}=\ket{5S_{1/2},F=1,m_F=0}$ and $\ket{d}=\ket{5S_{1/2},F=2,m_F=0}$ as well as the intermediate states $\ket{e_\pm}=\ket{5P_{3/2},F=1,m_F=\pm 1}$, with decay rate $\gamma/2\pi\sim\SI{6}{MHz}$, are suitable. The circular polarizations of the probe field and the counter-propagating control fields, would need to be chosen accordingly. The coupling to a Rydberg state $\ket{r}=\ket{nS_{1/2},J=1/2,m_J=1/2}$ is possible with a far detuned, two-photon transition, e.g. via an intermediate state in the $5P_{3/2},F=3$ hyperfine manifold. Coupling strengths used for the control and Rydberg coupling fields in Fig.~\ref{fig:transmission}, especially for small ratios of $\Omega_s/\gamma$, are experimentally accessible.

A quasi-1D situation may be achieved by trapping the atoms in a tightly confined optical dipole trap with a transversal waist, that is in the order of typical blockade radii of $5$ to $\SI{10}{\mu m}$, and an extend in propagation of about $\SI{40}{\mu m}$. An atomic density in the order of $\SI{1}{\mu m^{-3}}$ provides the optical depth used for the plots in Fig.~\ref{fig:transmission}.


%

\end{document}